# New Insight into the Properties of Proton Conducting Oxides from Neutron Total Scattering

Lorenzo Malavasi,*,[a] Hyunjeong Kim[b], Thomas Proffen[b], Giorgio Flor[a]

((Dedication, optional))

In recent years there has been a growing interest in searching for new proton conducting materials that could be successfully used in medium temperature solid oxide fuel cells (SOFC). In particular, proton conducting oxides have been the subject of a massive research activity[1-3]. Among the most promising oxide the acceptor doped cerates appears to be those most appealing in view of practical applications[4-5]. A relevant aspect of these materials is the investigation of the local distortion of the structure arising from water incorporation. This kind of study is of great help in defining how the structure changes in order to accommodate the proton which is usually thought to enter the structure in form of hydroxyl group according to the following equilibrium:

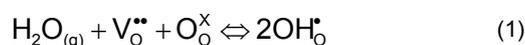

$$H_2O_{(g)} + V_O^{\bullet\bullet} + O_O^X \Leftrightarrow 2OH_O^{\bullet} \qquad (1)$$

where the oxygen vacancy results from the acceptor doping on the Ce site. Atomistic simulation work confirmed that the preferential location of dopant ions is on the Ce site[6].

To the best of our knowledge the only experimental work addressing the role of dopant and water incorporation on the local structure of Y-doped cerates is a X-ray absorption spectroscopy (XAS) work carried out by Longo and co-workers[7] at the Y K-edge. The main conclusion of that work was the observation that Y-doping induces a distortion of the parent $BaCeO_3$ structure resulting in a significantly distorted Y local environment. However, local structure information derived from XAS study does not provide a direct structural information and depends strongly upon the model used to calculate theoretical χ(k) which is not unique. Moreover, the XAS analysis usually provide significant information only up to the second shell. As a consequence, a more reliable and useful technique to investigate the local arrangement in these proton conducting oxides appears to be the Pair Distribution Function (PDF) analysis derived from total neutron scattering measurements (see the Experimental part for details).

The power of the PDF analysis in obtaining information on the short range order in complex materials is well proven.[8,9] The strength of the technique comes from the fact that it takes all the components of the diffraction data (Bragg peaks and diffuse scattering) into account and thus reveals both the longer range atomic order and the local deviations from it. In addition, its application to investigate the local order in solid state ionics materials has proved to be successful in unveiling new information on the short range order[10].

In the present work we investigated the pure $BaCeO_3$ and the acceptor doped $BaCe_{0.90}Y_{0.10}O_{2.95}$ compounds. In both cases the samples have been measured at room temperature and after being exposed to dry and wet air (humidification attained through bubbling air in $D_2O$). Aim of this work is to look at the effect of Y-doping and water doping on the local structure of the above mentioned samples.

Figure 1 shows the PDF of $BaCeO_3$ treated in dry (BCODRY) and wet air (BCOWET) up to 10 Å. As can be appreciated both PDF looks very similar without any significant deviations when treating the sample under wet conditions. First peak in the PDF around 2.4 Å is due to Ce-O pairs, the second one around 2.70 Å to Ba-O pairs, and the third one around 3.15 Å to O-O pairs.

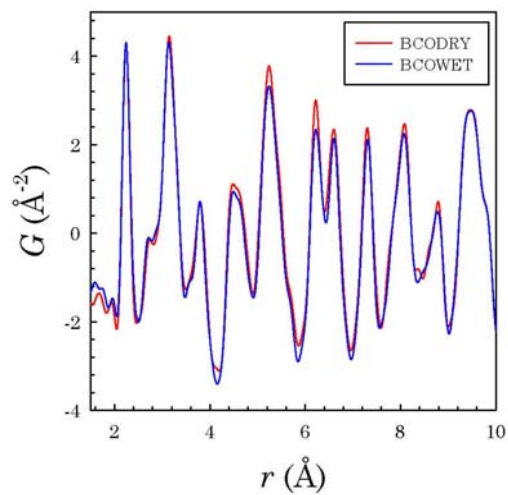

**Figure 1.** Comparison between the PDF of $BaCeO_3$ treated in dry (BCODRY) and wet air (BCOWET) up to 10 Å

Refinement of the PDF have been carried out starting from the model obtained with the Rietveld refinement of the average structure with the orthorhombic *Pmcn* space group. The fit result for the BCODRY sample is reported in Figure 2. As can be seen a very good refinement is obtained according to the available structure for BaCeO$_3$ obtained from diffraction data[11-12]. Refined parameters were: lattice constants, atomic positions, correlation parameters (which takes into account the correlated motion of atomic pairs at low-$r$) and isotropic atomic displacement parameters for the cations and anisotropic for the anions. The final agreement factor ($R_w$) was 7.9% which is a very good value for a PDF fit. The same model is able to nicely describe the local structure of BCOWET with a final $R_w$ of 8.3%.

[a]  Dr. L. Malavasi, Prof. G. Flor
     Department of Physical Chemistry "M. Rolla", INSTM and IENI-CNR
     University of Pavia
     Viale Taramelli 16, 27100 Pavia, Italy
     Fax: (+)39 382 987575
     E-mail: Lorenzo.malavasi@unipv.it
[b]  Dr. H. Kim, Dr. Th. Proffen
     Los Alamos National Laboratory
     LANSCE-12, MS H805, Los Alamos, NM

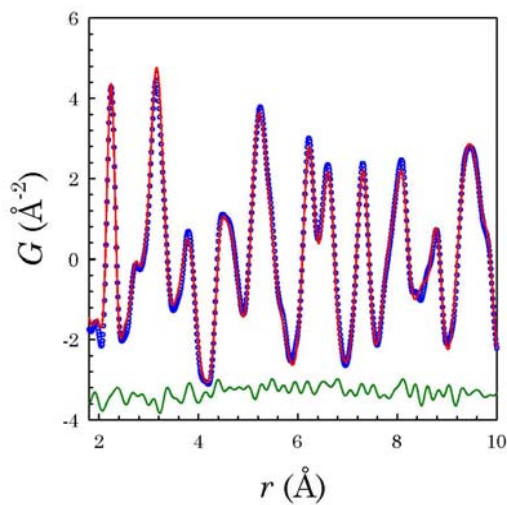

*Figure 2*. Fit of BCODRY PDF with *Pmcn* model. Blue circles represents the experimental G, red line the calculated G and the green line the difference.

Let us now pass to consider the Y-doped BaCeO$_3$. Figure 3 shows the PDF of the dry and wet BaCe$_{0.90}$Y$_{0.10}$O$_{2.95}$ samples (hereinafter called BCYDRY and BCYWET).

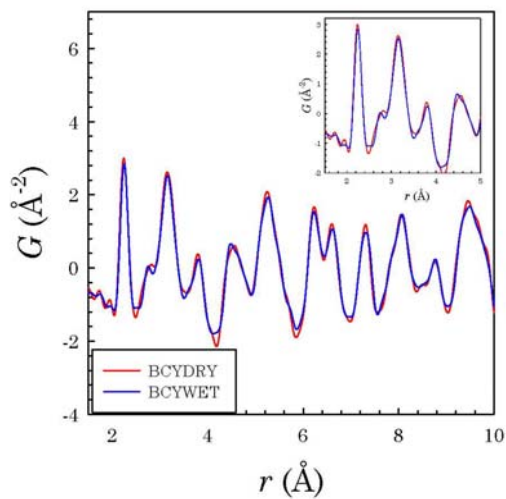

*Figure 3.* Comparison between the PDF of BaCe$_{0.9}$Y$_{0.1}$O$_{2.95}$ treated in dry (BCYDRY) and wet air (BCYWET) up to 10 Å.

First of all we may note that for BaCe$_{0.90}$Y$_{0.10}$O$_{2.95}$ the PDF of the two samples (dry and wet) are different. This correlates to the fact that the amount of absorbed water is in this case significant, due to the large concentration of oxygen vacancies introduced with the acceptor doping. For pure BaCeO$_3$ we did not expect to observe evident differences between the two samples – as experimentally confirmed - due to the unavailability of proton acceptor sites (oxygen vacancies).

A common feature of the short range structure of both BCY samples with respect to the BCO samples is a significant broadening of the first peak (Ce-O). For example, the FWHM of the this peak changes from ~0.142(3) for the BCODRY to ~0.163(3) for the BCYDRY sample. Qualitatively this broadening may be correlated to the presence of two cations on the B-site of the structure, where, for BaCeO$_3$, the only Ce ion was present. Since there are noteworthy differences between BCYDRY and BCYWET we will first discuss them separately. Together with Ce/Y-O peak broadening another relevant change in the PDF of BCYDRY samples is related to the Ba-O peak (see inset of Figure 3). Its shape and width is deeply influenced by the acceptor doping. While the average structure may still be refined by considering the orthorhombic space group the local environment deviates from it. The use of the *Pmcn* structure to model the PDF (even with anisotropic a.d.p. for all the ions) leads to unsatisfactorily results particularly for the short range order. The result is shown in Figure 4.

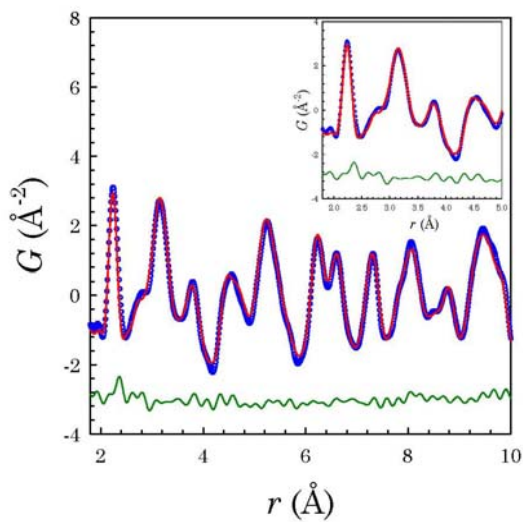

*Figure 4*. Fit of BCYDRY PDF with *Pmcn* model. Blue circles represents the experimental G, red line the calculated G and the green line the difference.

The overall agreement factor is not bad since already above ca. 6 Å the orthorhombic model seems to work well. However it is clear that the first coordination shells are poorly reproduce within this model compared to its application to describe the BaCeO$_3$ samples. In particular the worst agreement is found in the region between 2.5 and 3.5 Å. Many attempts were tried in order to develop a suitable model to describe the local structure of dry BaCe$_{0.90}$Y$_{0.10}$O$_{2.95}$. The simple inclusion of two different sets of bond lengths for the Ce/Y cations did not improve the fit. Our alternative approach was to try to use a space group of lower symmetry. We started the modelling by considering the sub-groups of the *Pmcn* space group. By none of these we could observe a significant improvement of the final agreement between observed and calculated PDF. The best result was achieved by considering a space group of relatively low-symmetry (*P2$_1$* - no. 4). In this space group all the atomic positions have a multiplicity of 2 and are in a general position ($x,y,z$). However, we maintained the special positions of the *Pmcn* space group for the Ba, Ce/Y and O atoms and $\beta$=90°. The result of the fit considering this lower-symmetry structure is reported in Figure 5.

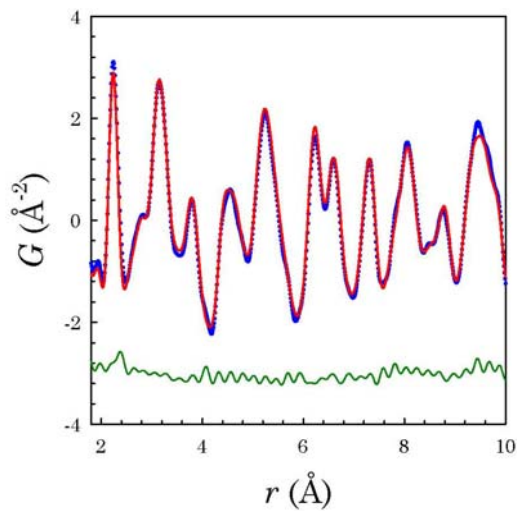

*Figure 5.* Fit of BCYDRY PDF with $P2_1$ model. Blue circles represents the experimental G, red line the calculated G and the green line the difference.

The improvement of the fit is now significant, particularly for the region that was poorly described in the *Pmcn* space group ($R_w$ pass from 12.7% in the *Pmcn* SG to 11.3% in the $P2_1$ SG). This structure leads to an increase in the octahedra distortion and in the general distortion of the Ba-O environment and to a greater distribution in the cation-oxygen bond lengths. Isotropic a.d.p. ($B$) for O1 and O2 are 0.76 and 1.16 Å$^2$. The refinement of the oxygen occupancies confirms some computational and experimental results[6,12] which indicate that the preferential site for their localization is the O2 site. In the present case the O1 site resulted to be fully occupied while the O2 site has an occupancy of 0.92(5). We believe that much of the distortion introduced within the lattice of BCYDRY arises from the oxygen vacancies created by the acceptor doping. However the structure accommodates this relatively high level of defects by significantly distorting on the local scale. When water is incorporated within the structure the short range order changes. This can be already appreciated by looking at Figure 3 where the PDF of BCYDRY and BCYWET were compared. A good refinement of the PDF of the BCYWET sample up to 10 Å can be again obtained by means of the orthorhombic *Pmcn* space group ($R_w$=9.0%). In particular, the region where this model markedly failed in the BCYDRY sample (2.5-3.5 Å) can be now nicely described within the orthorhombic structure. The fit result is displayed in Figure 6. Refinement of O occupancy leads in this case to both O sites fully occupied within the experimental error. Also for this fit all the a.d.p. display reasonable values, with O1 and O2 $B$-values of 0.8(1) and 1.2(1) Å$^2$.

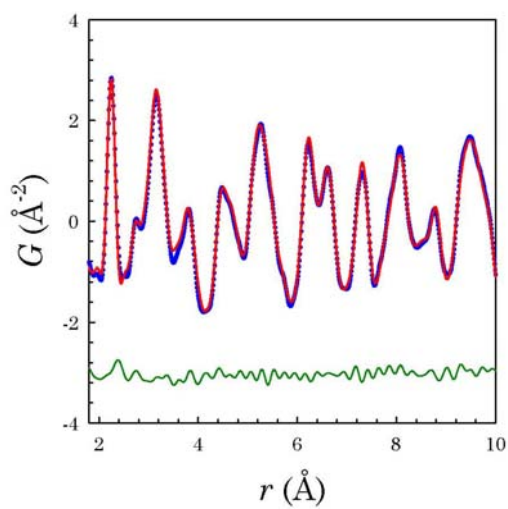

*Figure 6.* Fit of BCYWET PDF with *Pmcn* model. Blue circles represents the experimental G, red line the calculated G and the green line the difference.

The data acquired in the present total neutron scattering study revealed new and important features of proton conducting cerates. First of all we remark that this work represents the first PDF investigation of such class of materials where the application of PDF may shed light on aspects of the local order which are or relevance for the functional properties.

Firstly, we could observe that pure $BaCeO_3$ can be described in the short-range according to the average structure determined from diffraction. Protonation of this compound does not lead to any detectable influence on the material structure, thus confirming that the absence of oxygen vacancies do not permit any relevant hydration of the structure. Y-doping has a quite important effect on the short-range order of the unprotonated material. While the average structure has been nicely described within the same space group of the undoped sample (*Pmcn*)[11,12], the introduction of a foreign atom seems to distort the local environment around the cations with a significant effect on the Ba-O shell. The description of the short-range order requires to remove the symmetry of the *Pmcn* oxygen sites. Since Y and Ce have very similar atomic radii (1.04 and 1.01 Å, respectively, for the same coordination) the source of local distortion is most probably related to the creation of oxygen vacancies when Y is introduced in the lattice. When these oxygen vacancies are "filled" with water according to eq. (1), the structure locally orders and those sites which were crystallographically distinct in the dry sample are no longer so and the short-range order can be again described by means of the *Pmcn* structure. From these results we may observe that a requisite for a material to be a good proton conductor seems to be its ability to be flexible to distort when oxygen vacancies are created, thus allowing an easy water incorporation which restores the "original" symmetry which is more suited for proton conductivity through a Grotthus mechanism. This conclusion is partially supported by the observation that Sr-based cerates are poorer proton and oxygen conductors with respect to Ba containing cerates[13]. Ba is a "softer" ion with respect to Sr and forms less ionic and rigid bonds that allow the structure to deform more easily as required in the process of Protonation, as we have determined with our work. Of course also the activation energies for proton migration are another essential phenomenon in determining the proton conductivity and these can be related to the structural evidence here reported.

## Experimental Section

Powder samples of $BaCeO_3$ and of $BaCe_{0.9}Y_{0.1}O_3$ have been prepared by conventional solid state reaction from the proper stoichiometric amounts of BaO, $CeO_2$, and $Y_2O_3$ (all Aldrich ≥ 99,9%) by two successive firings for 20 h at 1200 °C[12]. X-ray diffraction analysis collected on a Bruker D8 Advance diffractometer (Cu radiation, 2θ-range 10-110°, step size 0.02°, time per step 5 sec) was used for assessment of phase purity of the as synthesised material. No traces of spurious phases have been detected. Half of the as-synthesised batch was annealed in dry air at 600°C and slowly cooled down to RT at 5°C/min while the other half was annealed in the same gas but after passing it trough a gas bubbler filled with $D_2O$.

Neutron powder diffraction measurements at RT were carried out on the NPDF diffractometer at the Lujan Center at Los Alamos

National Laboratory in a cylindrical vanadium tube. Data were processed to obtain the PDFs using PDFgetN[14] and the structural modelling was carried out using the program PDFFIT2[15].

*Acknowledgements*

*This work has been supported by the "Celle a combustibile ad elettroliti polimerici e ceramici: dimostrazione di sistemi e sviluppo di nuovi materiali" FISR Project of Italian MIUR. We recognize the support of the UNIPV-Regione Lombardia Project on Material Science and Biomedicine. This work has benefited from the use of NPDF at the Lujan Center at Los Alamos Neutron Science Center, funded by DOE Office of Basic Energy Sciences. Los Alamos National Laboratory is operated by Los Alamos National Security LLC under DOE Contract DE-AC52-06NA25396. The upgrade of NPDF has been funded by NSF through grant DMR 00-76488.*

**Keywords:** proton conductors, pair distribution function analysis, neutron scattering

[1] T. Norby, *Solid State Ionics* **1999**, *125*, 1.
[2] T. Norby, M. Widerøe, R. Glöckner and Y. Larring, *Dalton Trans.* **2004**, *19*, 3012-3018.
[3] K.D. Kreuer, *Solid State Ionics* **1997**, *97*, 1.
[4] H. Iwahara. T. Esaka, H. Uccida, and N. Maeda, *Solid State Ionics* **1981**, *3–4*, 359.
[5] H. Iwahara, H. Uchida, K. Ono, J. Ogaki. *J. Electrochem. Soc.* **1988**, *135*, 529.
[6] M.S. Islam, R.A. Davies, J.D. Gale, *Chem. Mater.* **2001**, *13*, 2049-2055.
[7] Longo, A.; Giannici, F.; Balerna, A.; Ingrao, C.; Deganello, F.; Martorana, A. *Chem. Mater.* **2006**, *18*, 5782.
[8] Billinge, S. J. L. & Kanatzidis, M. G., *Chem. Commun.* 749-760 (2004).
[9] Egami, T. & Billinge, S. J. L. *Underneath the Bragg peaks: structural analysis of complex materials* (Pergamon Press, Elsevier, Oxford, England, 2003).
[10] L. Malavasi, H. Kim, Simon J. L. Billinge, Th. Proffen, C. Tealdi, G. Flor, *J. Am. Chem. Soc.* **2007**, *129*, 6903.
[11] K.S. Knight, *Solid State Ionics* **2001**, *145*, 275.
[12] L. Malavasi, C. Ritter, G. Chiodelli, *Chem. Mater.* **2008**, *20*, 2343.
[13] G.C. Mather and M.S. Islam, *Chem. Mater.* **2005**, *17*, 173.
[14] P.F. Peterson, M. Gutmann, Th. Proffen, S.J.L. Billinge, *J. Appl. Crystallogr.* **2000**, *33*, 1192.
[15] C. L. Farrow, P. Juhas, J. W. Liu, D. Bryndin, E. S. Bozin, J. Bloch, Th. Proffen and S. J. L. Billinge, *J. Phys.: Condens. Matter* **2007**, *19*, 335219.

**Entry for the Table of Contents** (Please choose one layout)

Layout 1:

# COMMUNICATIONS

In this paper we investigated the most important family of proton conducting oxides, i.e. cerates, by means of pair distribution function analysis (PDF) obtained from total neutron scattering data.

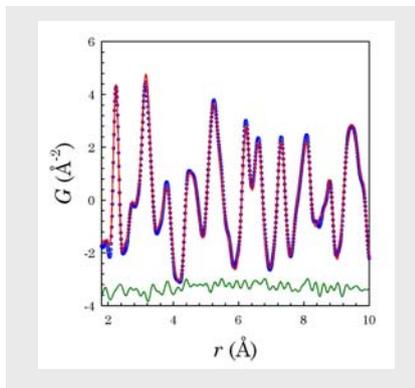

*Lorenzo Malavasi,\*, Hyunjeong Kim, Thomas Proffen, Giorgio Flor*

**Page No. – Page No.**

**((Title))**